\documentclass[aps,prl,groupedaddress,reprint]{revtex4-1}

\usepackage{float}
\usepackage{graphics}
\usepackage{graphicx}
\usepackage{epsfig}
\usepackage{epstopdf}

\newcommand{\bea}{\begin{eqnarray}}
\newcommand{\eea}{\end{eqnarray}}
\newcommand{\be}{\begin{equation}}
\newcommand{\ee}{\end{equation}}
\newcommand{\mw}   {\mbox{$M_W$}}

\newcommand{\mt}   {\mbox{$m_t$}}
\newcommand{\mz}   {\mbox{$M_Z$}}
\newcommand{\mzsq}   {M_Z^2}

\newcommand{\sefff}{\sin^2\theta_{eff}^{f}\,}
\newcommand{\seffl}{\sin^2\theta_{eff}^{\ell}\,}

\newcommand{\afb}   {A_{FB}}

\begin{document}

\preprint{CERN-TH-2019-092}

\title{On the direct determination of $\sin^2\vartheta_{eff}^\ell$ at hadron colliders}



\author{Mauro Chiesa}
\email[]{Mauro.Chiesa@physik.uni-wuerzburg.de}
\affiliation{Institut f\"ur Theoretische Physik und Astrophysik,
  Julius-Maximilians-Universit\"at W\"urzburg,
Emil-Hilb-Weg 22, D-97074 W\"urzburg, Germany }

\author{Fulvio Piccinini}
\email[]{fulvio.piccinini@pv.infn.it}
\affiliation{INFN, Sezione di Pavia, Via A. Bassi 6, 27100 Pavia, Italy}

\author{Alessandro Vicini}
\email[]{alessandro.vicini@mi.infn.it}
\affiliation{TH Department, CERN 1 Esplanade des Particules, Geneva 23, CH-1211, Switzerland and\\
  Dipartimento di Fisica ``Aldo Pontremoli'', University of Milano and INFN Sezione di Milano,
  Via Celoria 16, 20133 Milano, Italy}


\date{\today}

\begin{abstract}
  We discuss the renormalization of the electroweak Standard Model at 1-loop
  using the leptonic effective weak mixing angle as one of the input parameters.
  We evaluate the impact of this choice in the prediction of the forward-backward asymmetry
  for the neutral current Drell-Yan process. 
  The proposed input scheme is suitable for a direct determination of the effective
  leptonic weak mixing angle from the experimental data.
\end{abstract}


\maketitle


\section{Introduction}

The weak mixing angle \cite{Glashow:1961tr,Weinberg:1967tq,Salam:1968rm,Sirlin:1980nh}
is a fundamental parameter of the theory of the electroweak (EW) interaction,
as it determines the combination of the gauge fields
associated to the third component of the weak isospin and to the hypercharge,
yielding the photon and the $Z$ boson fields.
The leptonic effective weak mixing angle $\seffl$, defined at the $Z$ resonance,
has been proposed 
  \cite{Marciano:1980pb,Degrassi:1990ec,Gambino:1993dd,Bardin:1997xq,Arbuzov:2005ma,Degrassi:2014sxa}
as a quantity sensitive to new physics,
offering the opportunity of a stringent test of the Standard Model (SM).
The measurement at LEP/SLD \cite{ALEPH:2005ab} \footnote{
The measurement of $\seffl$ was based on the parameterisation of the $Z$ resonance in terms of pseudo-observables.
The experimental values of the latter were then fitted with tree-level expressions of the initial- and final-state currents,
with the weak mixing angle as a free parameter which was then interpreted as an effective quantity. 
  }
has been later challenged by the CDF and D0 determinations \cite{Aaltonen:2018dxj} at the Fermilab Tevatron and more recently by the results from the LHC collaborations
ATLAS \cite{ATLAS:2018gqq},
CMS \cite{Sirunyan:2018swq}
and LHCb \cite{Aaij:2015lka}.
Two conceptually different strategies can (and should) be pursued
for the direct determination of $\seffl$:
with,
whenever possible,
a model independent as well
as a pure SM approach. The latter will be useful as an internal self-consistency
check of the SM, through the comparison of the direct determination with the
most precise available calculations of $\seffl$. 
In this paper we discuss the renormalization of the EW SM at 1-loop level,
using $\seffl$,
as defined at LEP/SLD,
as one of the input parameters in the EW gauge sector. 
Any simulation code implementing such a scheme
will be able to provide theoretical templates
for a direct sensible comparison with the experimental data,
with the leptonic effective weak mixing angle used as a fit parameter and consistently treated
in the evaluation of NLO and higher-order corrections.
The use of $\seffl$ as input parameter of the electroweak sector
has also been proposed in Refs.~\cite{Kennedy:1988sn,Renard:1994ay,Ferroglia:2001cr,Ferroglia:2001ps,Ferroglia:2002rg}
in the framework of the high-precision measurements at the $Z$ boson resonance and  higher energies 
at future $e^+e^-$ colliders.

\section{\label{sec:inputrenormalization} Input schemes and renormalization}
The choice of an input scheme in the EW gauge sector of the SM is relevant for two distinct reasons:
\begin{enumerate}
  \item
In a theoretical perspective,
the prediction of an observable should be affected
by the smallest possible parametric uncertainty.
This goal can be achieved by using the best known measured constants,
like for instance the fine structure constant $\alpha$, 
the Fermi constant $G_\mu$ and the $Z$ boson mass $\mz$. 
Furthermore, the convergence of the perturbative expansion used to predict
an observable is an additional criterium to judge whether the chosen inputs 
describe the process already in lowest order with good accuracy 
and reabsorb in their definition large radiative corrections.
This is the case, for instance, of the scheme which uses $G_\mu$, $\mz$ 
and the $W$ boson mass $\mw$, to describe processes 
at the electroweak and higher scales.

  \item
    The determination of a fundamental constant at high-energy colliders
    can be achieved through the comparison of kinematical distributions 
    computed in a theoretical model, the so called templates,
    with the experimental data.
    The fundamental constant must be a free parameter of the model 
    and is varied in the fitting procedure.
    Only the input parameters of the model can be unambiguously determined, 
    because they are the only ones which can be freely varied without spoiling the
    accuracy of the calculation, 
    while any other quantity is a prediction expressed in terms of them.
    Typical examples have been $\mz$ at LEP1 and $\mw$ at LEP2, Tevatron and LHC.
    
\end{enumerate}
Following the second perspective,
we discuss in this paper the formulation of a renormalization scheme
which includes the leptonic effective weak mixing angle $\seffl$ \cite{Marciano:1980pb}
as one of the input parameters. Such a scheme will allow to exploit
the Tevatron and LHC (and in particular the future HL-LHC) potential
to provide very high precision measurements
of the neutral channel (NC) Drell-Yan (DY) process and, in turn, of $\seffl$.

\subsection{Input scheme definitions}
A set of three commonly adopted SM lagrangian input parameters in the gauge
sector is $e, M_W, M_Z$;
they have to be expressed in terms of three measured quantities,
whose choice defines a renormalization scheme. 
The relation between $e, M_W, M_Z$ and the reference measured quantities 
has to be evaluated at the same perturbative order of the scattering amplitude
calculation at hand and allows to fix the renormalization conditions. 
The usual sets of reference measured quantities are:
$\alpha, M_W, M_Z$, which defines the on-shell scheme;
$\alpha(M_Z), M_W, M_Z$, which is a variant of the on-shell
scheme and reabsorbs the large logarithmic contributions due to the running of
the electromagnetic coupling from the scale $0$ to $M_Z$~\footnote{The uncertainties related to the hadronic contribution to the running of the QED coupling constant can be evaluated through dispersive relations based on $e^+ e^-$ data at low energies~\cite{Jegerlehner:2018zrj,Keshavarzi:2018mgv,Davier:2017zfy}}; $G_\mu, M_W, M_Z$,
which defines the $G_\mu$ scheme and is particularly suited to describe DY
processes at hadron colliders because it allows to include a large part of the radiative corrections
in the LO predictions, guaranteeing a good convergence of the perturbative series. 
For a detailed description of these schemes cfr. Ref.~\cite{Dittmaier:2009cr}.
The presence of $M_W$ among the input parameters is a nice feature in view of
a direct $M_W$ determination at hadron colliders via a template fit method, as described above.
On the other hand, these schemes are not suited for high precision predictions,
because of the ``large'' parametric uncertainties stemming from the present experimental
precision on the knowledge of $M_W$. In fact, for NC DY precise predictions,
a LEP style scheme with $\alpha, G_\mu, M_Z$ would be preferred. However, in view of a direct SM determination 
of the quantity $\seffl$, also this scheme has its own shortcomings, because $\seffl$
is a calculated quantity and can not be treated as a fit parameter.
With the aim of a direct $\seffl$ SM determination, 
we discuss an alternative scheme, which includes the weak 
mixing angle as a SM lagrangian input parameter, $\sin^2\theta$, together with $e$ and $M_Z$. The experimental
reference data are the $Z$ boson mass value measured at LEP, the
fine structure constant $\alpha$ and $\seffl$ as defined at LEP at the $Z$ resonance.
An additional possibility discussed in the following is to replace $\alpha$
with $G_\mu$. 
We will refer to these two choices as the 
$(\alpha,\mz,\seffl)$ and the $(G_\mu,\mz,\seffl)$ input schemes.
At tree level $\sin^2\theta = \seffl$. 
The quantity $\seffl$ is defined in terms of the vector and axial-vector couplings
of the $Z$ boson to leptons $g_{V,A}^\ell$, measured at the $Z$ boson peak,
or alternatively the chiral electroweak couplings $g_{L,R}^\ell$ and reads
(at tree level)~\footnote{Analogous relations can be written for different
fermion species $f$, yielding flavour dependence of $\sefff$ beyond tree level.}:
\be
\sin^2\theta=
\sin^2\theta_{eff}^\ell 
=\frac{I_3^{l}}{2 Q_l}\left(1-\frac{g_V^\ell}{g_A^\ell} \right)
=\frac{I_3^{l}}{Q_l}\left(\frac{-g_R^\ell}{g_L^\ell-g_R^\ell} \right)\, ,
\label{eq:sinthetatree}
\ee
where
\be
g_L^\ell = \frac{I_3^{l}-\sin^2\theta_{eff}^\ell\, Q_l}{\sin\theta_{eff}^\ell \cos\theta_{eff}^\ell}\, ,\quad    \quad g_R^\ell ~=~-\frac{\sin\theta_{eff}^\ell}{\cos\theta_{eff}^\ell} Q_l\, .
\ee
$I_3^{l}= - \frac12$ is the third component of the weak isospin
and $Q_l$ is the electric charge of the lepton in units of the positron charge.

\subsection{Renormalization}

We implement the one loop renormalization of the three input parameters
by splitting the bare ones into renormalized parameters and counterterms
\bea
M_{Z,0}^2 &=& \mzsq + \delta\mzsq
\label{eq:ctmz}\\
\sin^2\theta_0&=&\seffl + \delta\seffl
\label{eq:ctsin}\\
e_0 &=& e (1 + \delta Z_e)
\label{eq:cte}
\eea
where the bare parameters are denoted with subscript $0$. 
The counterterms $\delta\mzsq$ and $\delta Z_e$ are defined as in the usual on-shell
scheme. Complete expressions are given in Eqs.~(3.19) and (3.32) of 
Ref.~\cite{Denner:1991kt}.
The counterterm $\delta\seffl$ is defined by imposing that the tree-level
relation Eq.~(\ref{eq:sinthetatree}) holds to all orders. 
Considering the $Z \ell^+ \ell^-$ vertex and neglecting the masses of the lepton $\ell$,
the couplings $g_{L,R}^\ell$ are replaced by the form factors 
${\cal G}_{L,R}^\ell(q^2)$~\cite{Arbuzov:2005ma}
once radiative corrections are accounted for. 
The effective weak mixing angle has been defined at LEP/SLD 
by taking the form factors
at $q^2=\mzsq$:
\be
\seffl
~\equiv~\frac{I_3^{l}}{Q_l}\,{\rm Re}\left(\frac{-{\cal G}_R^\ell(\mzsq)}{{\cal G}_L^\ell(\mzsq)-{\cal G}_R^\ell(\mzsq)} \right)\, .
\label{eq:def-sinthetaeff}
\ee

The form factors ${\cal G}_i^\ell$ can be computed in the SM
in any input scheme that does not contain $\seffl$ as input parameter,
yielding in turn, via Eq.~(\ref{eq:def-sinthetaeff}), a prediction for $\seffl$,
as discussed at length in Refs.~\cite{Degrassi:1996ps,Awramik:2006uz}.

In this paper instead we consider the weak mixing angle as an input parameter.
In order to fix its renormalization condition,
we write Eq.~(\ref{eq:def-sinthetaeff}) at one-loop 
\be
\seffl~=~
\frac{I_3^{l}}{Q_l}\,
     {\rm Re}\left(
\frac{-g_R^\ell-\delta g_R^\ell}{g_L^\ell-g_R^\ell+\delta g_L^\ell -\delta g_R^\ell}
\right)\, ,
\label{eq:sinthetaren}
\ee
where $\delta g_{L,R}^\ell$ represent the effect of radiative corrections,
expressed in terms of renormalized quantities and related counterterms,
including $\delta\seffl$.
We do not consider NLO QED corrections because they factorize on form factors
and therefore do not affect the $\sin^2\theta_{eff}^\ell$ definition. 
The effective weak mixing angle is defined to all orders by the request that the measured value coincides with the
tree-level expression. The counterterm $\delta\seffl$ is fixed by imposing that the
one-loop corrections to Eq.~(\ref{eq:sinthetatree}) vanish, namely:
\be
\frac12\,
     \frac{g_L^\ell g_R^\ell}{(g_L^\ell-g_R^\ell)^2} \,
     {\rm Re} \left(
\frac{\delta g_L^\ell}{g_L^\ell}-
\frac{\delta g_R^\ell}{g_R^\ell}
     \right)\, ~=~ 0.
\label{eq:deltasintheta_0}
\ee
We remark that at one-loop the condition in Eq.~(\ref{eq:deltasintheta_0}) holds 
also if $\seffl$ is defined from the ratio of the real parts of ${\cal G}_V$ and ${\cal G}_A$.
Moreover, Eq.~(\ref{eq:deltasintheta_0}) remains unchanged if the complex-mass scheme~\cite{Denner:1999gp,Denner:2005fg,Denner:2006ic}
is used for the treatment of unstable particles. 
From the ${\cal O}(\alpha)$ corrections to the vertex $Z \ell^+\ell^-$
we obtain
\bea
&&\frac{\delta \seffl}{\seffl}
= {\rm Re} \Big\{ \frac{\cos \theta_{eff}^\ell}{\sin \theta_{eff}^\ell} \frac{\Sigma_T^{AZ}(M_Z^2)}{M_Z^2} \label{eq:deltasintheta_1} \\
                 && + 
                  \left( 1 - \frac{Q_\ell}{I_3^{\ell}} \seffl  \right)
                  \left[ \delta Z^\ell_L + \delta V^\ell
                    - \delta Z^\ell_R - \delta V^R\right] \Big\}. \nonumber
\eea
where $\Sigma_T^{AZ}(\mzsq)$ contains the fermionic and bosonic contributions 
to the $\gamma Z$ self-energy corrections, while the second line of
Eq.~(\ref{eq:deltasintheta_1}) stems from the vertex corrections and counterterm
contributions. We remark that the $\gamma Z$ self-energy does not contain enhanced
terms proportional to $m_t^2$. 
The bosonic contributions in Eq.~(\ref{eq:deltasintheta_1}) form a gauge invariant
set because they are a linear combination of the corrections to the left- and
right-handed components of the $Z$ decay amplitude into a lepton pair. 
The expression of $\Sigma_T^{AZ}(\mzsq)$ and $\delta Z_{L/R}^\ell$ are
given in Eqs.~(B.2) and (3.20) of Ref.~\cite{Denner:1991kt}, respectively. In
$\delta Z_{L/R}^\ell$ we suppressed the lepton family indices. The vertex corrections
$\delta V^{L/R}$ are given by
\bea
\delta V^\ell &=& \left( g^\ell_L \right)^2 \frac{\alpha}{4 \pi} {\cal V}_a\left( 0, M_Z^2, 0, M_Z, 0, 0\right) \nonumber \\
                 &+& \frac{1}{2 s_W^2} \frac{g^{\nu}_L}{g^{\ell}_L} \frac{\alpha}{4 \pi} {\cal V}_a\left( 0, M_Z^2, 0, M_W, 0, 0\right) \nonumber \\
     &-& \frac{c_W}{s_W}\frac{1}{2 s_W^2} \frac{1}{g^{\ell}_L} \frac{\alpha}{4 \pi} {\cal V}_b\left( 0, M_Z^2, 0, 0, M_W, M_W\right) \nonumber \\
\delta V^R &=& \left( g^\ell_R \right)^2 \frac{\alpha}{4 \pi} {\cal V}_a\left( 0, M_Z^2, 0, M_Z, 0, 0\right) 
\label{eq:vertex-v2}
\eea
and the vertex functions ${\cal V}_a$ and ${\cal V}_b$ are given in Eqs.~(C.1) and (C.2)
of Ref.~\cite{Denner:1991kt}, respectively.

The renormalization condition that
  the measured effective leptonic weak mixing angle matches the tree-level expression to all orders in perturbation theory
  applies, following the LEP definition, to the real part of the
  ratio of the vector and axial-vector form factors.
  The latter develop, order by order, an imaginary part which is computed
  in terms of the input parameters and contributes to the scattering amplitude.

\subsection{The $G_\mu$ scheme}
The muon decay amplitude allows to establish a relation between $\alpha, G_\mu, \mz$ and $\seffl$
which reads
\begin{equation}
  \seffl \cos\theta^2_{eff} M_Z^2 =
  \frac{\pi \alpha}{\sqrt{2}G_\mu} \left( 1+ \Delta \tilde r\right) \, .
  \label{eq:alpha-gmu_oneloop}
\end{equation}
with the following expression for $\Delta \tilde r$
\begin{eqnarray}
  \Delta\tilde r &=&  \Delta \alpha(s) - \Delta \rho + \Delta\tilde r_{rem}\label{eq:our-deltar_5}
  \\
\Delta\tilde r_{rem} &=&
  \frac{{\rm Re}\Sigma^{AA}(s)}{s} 
   - \left( \frac{\delta M_Z^2}{M_Z^2}
   - \frac{\Sigma_T^{ZZ}(0)}{M_Z^2} \right) \nonumber\\
   &+& \frac{s_W^2 - c_W^2}{c_W^2}\frac{\delta s_W^2}{s_W^2}  +2 \frac{c_W}{s_W}\frac{\Sigma_T^{AZ}(0)}{M_Z^2}    \nonumber\\
   &+& \frac{\alpha}{4\pi s_W^2} \Big( 6+ \frac{7-4s_W^2}{s_W^2} {\rm log} ( c_W^2 ) \Big) ,
\end{eqnarray}
where $s_W=\sin\theta_{eff}^{\ell}$ and $c_W=\cos\theta_{eff}^{\ell}$, respectively.  
We note the appearance of the combination $\Delta \alpha(s) - \Delta \rho$, which
differs from the corresponding one for $\Delta r$ in the $(\alpha,\mw\mz)$ on-shell scheme
$\Delta \alpha(s) - \frac{c_W^2}{s_W^2} \Delta \rho$.
The $\Delta\tilde r_{rem}$ correction does not contain any term enhanced by a $\mt^2$ factor,
nor large logarithmically enhanced contributions.
Using Eq.(~\ref{eq:alpha-gmu_oneloop}) to derive an effective electromagnetic coupling,
it is possible to convert results computed in  the $(\alpha,\mz,\seffl)$ scheme
in the corresponding ones in the $(G_\mu,\mz,\seffl)$ schemes.
The $\Delta\rho^{(1)}\equiv \Delta\rho$ term present at ${\cal O}(\alpha)$ in this relation
accounts for 1-loop quantum corrections growing like $\mt^2$;
the latter can be resummed to all orders,
together with the irreducible 2-loop contributions $\Delta\rho^{(2)}$,
computed in the heavy top limit in Ref.~\cite{Fleischer:1993ub}.
In the following predictions for the $(G_\mu,\mz,\seffl)$ scheme, we include
the effect of the universal $\mt^2$ corrections at two-loops with the
replacement $G_\mu \to G_\mu \left(1+\Delta\rho^{(1)}+\Delta\rho^{(2)}\right)$
after subtracting the $\Delta\rho^{(1)}$ contributions already included in
the one-loop calculation.

\section{The Drell-Yan process}
\label{sec:dy}
We study at NLO-EW the neutral current (NC) DY process,
in the setup described in \cite{Alioli:2016fum}
but without acceptance cuts on the lepton transverse momentum and pseudorapidity, with $\mw=80.385$ GeV, $\mt=173.5$ GeV and $\seffl=0.23147$.
The distributions are simulated
with the {\tt POWHEG} code ({\tt Z\_BMNNPV} processes svn revision 3652, under
the {\tt POWHEG-BOX-V2} framework)\cite{Barze:2013fru},
focusing on the lepton-pair invariant mass 
forward-backward asymmetry $\afb(M_{\ell^+ \ell^-}^2)$,
defined as $(F-B)/(F+B)$,
where $F=\int_0^1 dc\, d\sigma/dc$ and $B=\int_{-1}^0 dc\, d\sigma/dc$,
for a given value of $M_{\ell^+\ell^-}^2$
with $c$ the cosinus of the scattering angle in the Collins-Soper frame.
Given the gauge invariant separation of photonic and weak corrections,
we focus on the latter to discuss the main features of the $(G_\mu,\mz,\seffl)$ schemes,
in view of a direct determination of $\seffl$.

\begin{figure}
  \includegraphics[width=0.48\textwidth]{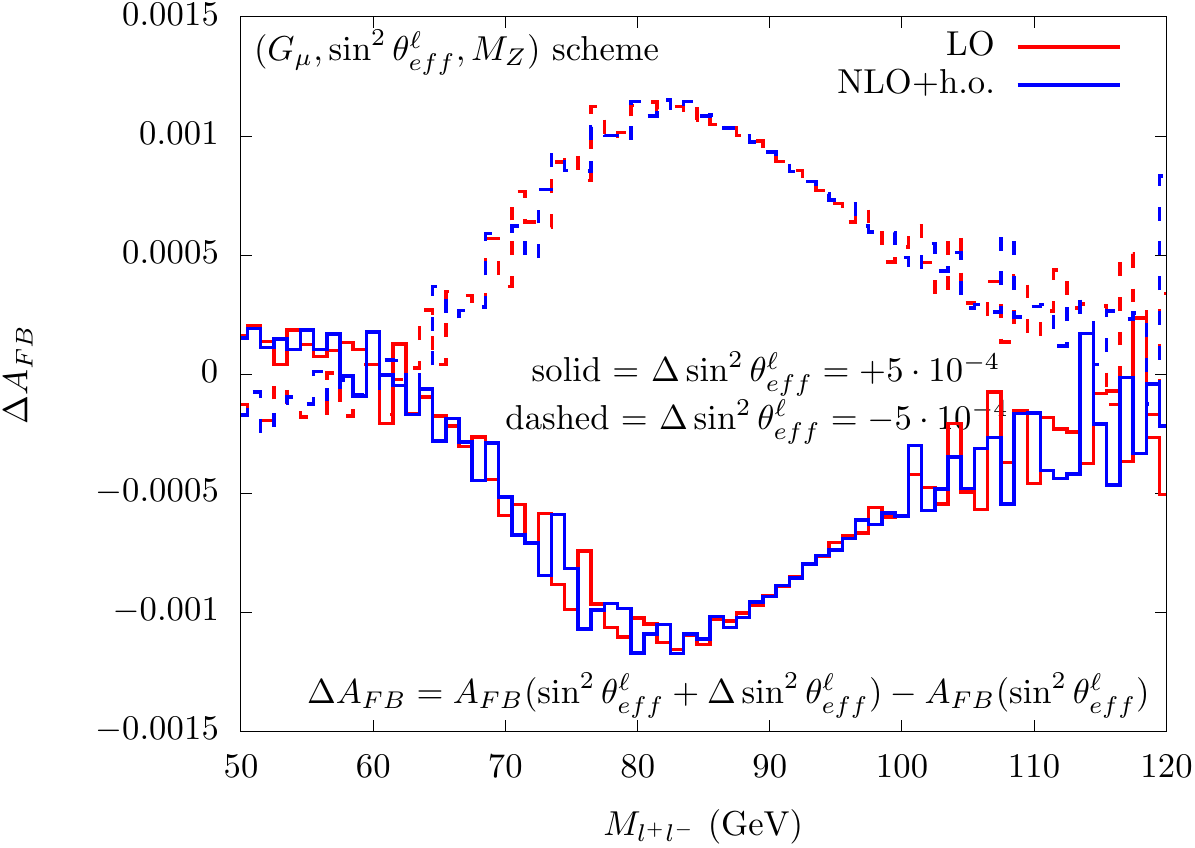}
  \caption{The absolute variation of the predictions for the forward-backward asymmetry
    by changing $\seffl$ by
    $\Delta = \pm 5\cdot 10^{-4}  $ w.r.t. the value 0.23147d0,
    using the $G_\mu$, $\seffl$, $M_Z$ scheme, at NLO-ho and LO
    accuracies. \label{fig:A_FB_gmu_sin2var5}}
\end{figure}

The absolute change of $\afb$ computed with two $\seffl$ values differing by $\Delta\seffl=5\cdot 10^{-4}$,
for a fixed choice of all the other inputs,
is shown in Fig.~\ref{fig:A_FB_gmu_sin2var5}.
The observed $\afb$ shift sets the precision goal of a measurement that aims at the determination of $\seffl$
at the level of $\Delta\seffl$.
Taking as a reference $\Delta\seffl=1\cdot 10^{-4}$ as a final precision goal at the LHC,
the results of Fig.~\ref{fig:A_FB_gmu_sin2var5} must be rescaled, in first approximation, by a factor 5.

\begin{figure}
  \includegraphics[width=0.48\textwidth]{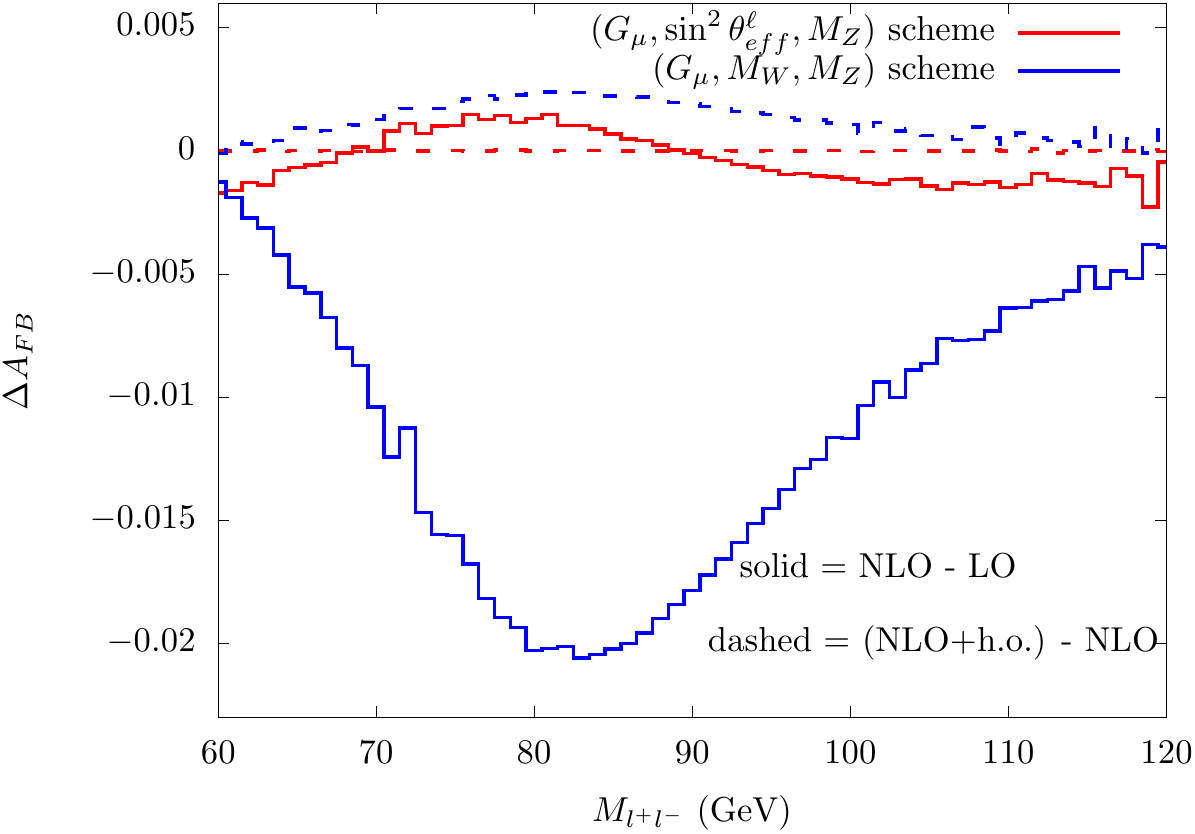}
  \caption{The absolute deviation of NLO(NLO-ho) w.r.t. LO(NLO) predictions on the lepton forward-backward asymmetry, in the renormalization scheme with 
  $G_\mu$, $\seffl$ as input. \label{fig:A_FB_sin2theta_gmu}}
\end{figure}

The absolute change $\Delta\afb$ of $\afb(\mzsq)$ computed with NLO weak virtual corrections
with respect to the LO result,
and the variation obtained with improved couplings with respect to the NLO case
are shown in Fig.~\ref{fig:A_FB_sin2theta_gmu} for the  $(G_\mu,\mz,\seffl)$ scheme (red lines)
and for the $(G_\mu,\mw, \mz)$  scheme (blue lines).
The comparison of the blue and red lines shows
a reduction by almost one order of magnitude in the $(G_\mu,\mz,\seffl)$ scheme
for the value of $\Delta\afb$ due to the inclusion of the NLO corrections;
we observe a negligible residual correction due to higher-order terms (h.o.),
at variance with the $(G_\mu,\mw, \mz)$ case, where we have a shift at the few parts $10^{-4}$ level
in the $Z$ peak region. The universal h.o. corrections in the $(G_\mu,\mw, \mz)$ scheme are
estimated according to Ref.~\cite{Dittmaier:2009cr}.

The size of NLO and higher-order radiative corrections, smaller than in the  $(G_\mu,\mw, \mz)$ case,
can be ascribed to the choice as input parameters of the quantities that parameterize the $Z$ resonance in terms of
normalization ($G_\mu$), position ($\mz$) and shape ($\seffl$), the latter two being defined at the $Z$ resonance
and thus reabsorbing a good fraction of the quantum corrections.

\begin{figure}
  \includegraphics[width=0.48\textwidth]{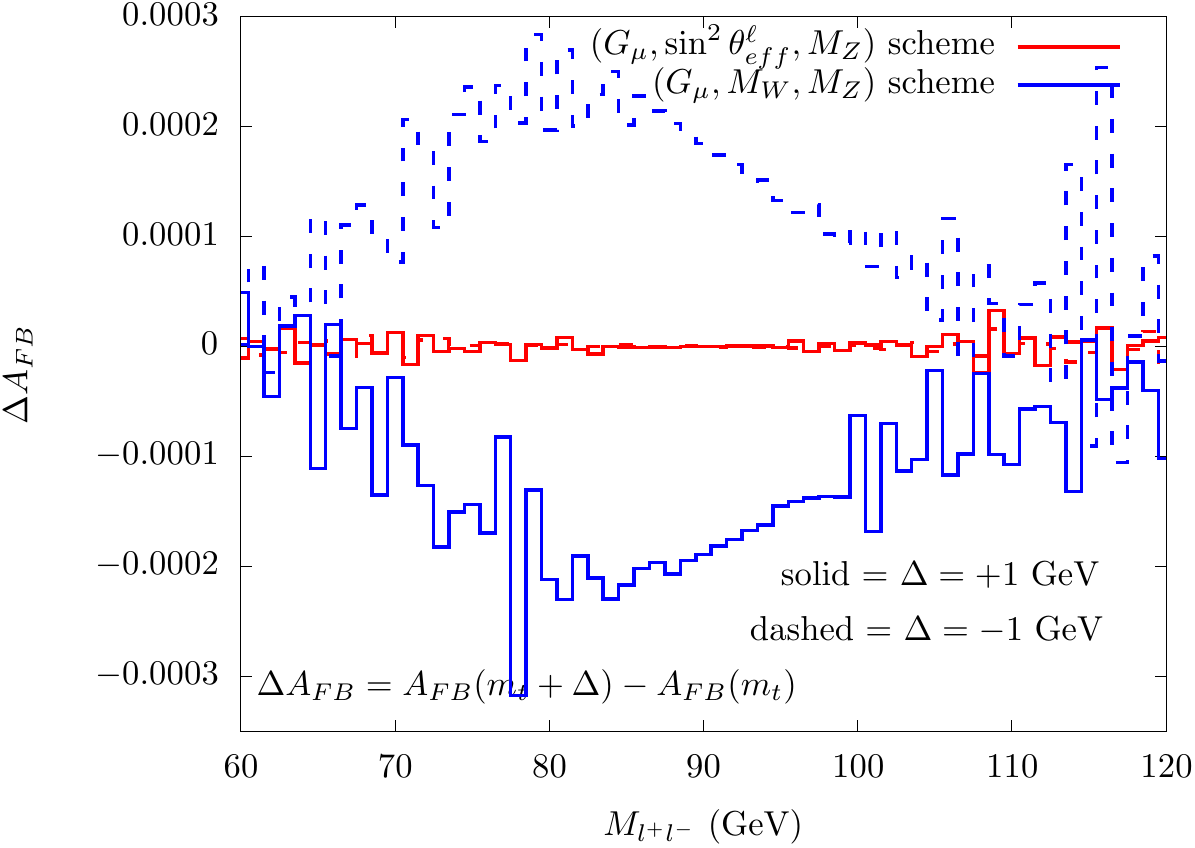}
  \caption{The absolute deviation between predictions on the lepton pair $A_{FB}$ as a function
    of $M_{\mu^+\mu^-}$, in the renormalization scheme with 
    $G_\mu$, $\seffl$ as input, with a variation of $\mt$ of
    $\pm$~1~GeV around the value $\mt = 173.5$~GeV. The precision of the
    calculation is NLO. \label{fig:A_FB_sin2theta_mt1}}
\end{figure}

One of the main sources of parametric uncertainties is given,
in any scheme with $G_\mu$ as input, by the value of $\mt$. 
In Fig.~\ref{fig:A_FB_sin2theta_mt1} 
we show the absolute variation of 
$\Delta A_{FB}$ w.r.t. a change of $\pm 1$~GeV of
$\mt$ around its central value, taken at $\mt = 173.5$~GeV,
using the NLO accuracy with higher order effects included,
evaluated in the $(G_\mu,\mz,\seffl)$ (red lines) and $(G_\mu,\mw, \mz)$ (blue lines) schemes. 
In the $(G_\mu,\mz,\seffl)$ scheme, 
the effect is well below the $2\cdot 10^{-5}$ scale for $\afb$ in the $[60,120]$ GeV mass range,
almost vanishing in the $Z$ peak region,
while in the $(G_\mu,\mw, \mz)$ case a variation of $\mt$ by $\pm 1$ GeV
induces a shift $\Delta\afb$ of order $2\cdot 10^{-4}$.
The very small dependence of $\afb$ on the $\mt$ value is due to
the cancellation of the overall normalization factor of the squared matrix element,
between numerator and denominator of $\afb$,
where the factor with the $\mt^2$ dependence is present.
Radiative corrections, logarithmic in $\mt$, are by construction small at the $Z$ peak,
so that also the residual $\mt$ dependence is milder than in other invariant mass regions.
In the $(G_\mu,\mw, \mz)$ case instead the $\mt^2$ dependence enters via the corrections to $\mw$
and affects the precise value of the on-shell weak mixing angle
and, in turn, the shape of the $\afb$ distribution.

In conclusion, we have presented an EW scheme that has $\seffl$,
with  exactly the same definition adopted at LEP/SLD,
among the input parameters of the gauge sector and
discussed its 1-loop renormalization. 
In such a scheme the predictions of the NC DY process
exhibit a faster convergence of the perturbative expansion
and smaller $\mt$ parametric uncertainties,
with respect to the $(G_\mu,\mw,\mz)$ scheme.
The presence of $\seffl$ among the inputs allows
its direct determination at hadron colliders
and
a closure test with a comparison against its best theoretical prediction
in the SM based on the $(\alpha,G_\mu,\mz)$ input scheme.
Such a scheme will allow the preparation of templates and the quantitative evaluation of
the impact of radiative corrections and other theoretical uncertainties,
in analogy to the study presented in Ref.~\cite{CarloniCalame:2016ouw} in the $\mw$ case.
We implemented the scheme in the {\tt Z\_BMNNPV} svn revision 3652 processes
under the 
{\tt POWHEG-BOX-V2} framework, but it can be easily adopted by any other code.

\begin{acknowledgments}
  We would like to thank D. Wackeroth, G. Degrassi, C.M. Carloni Calame,
  G. Montagna and O. Nicrosini for useful discussions and
  a careful reading of the manuscript.
  We would also like to thank  S. Dittmaier  and all colleagues
  of the LHC Electroweak Working Group for useful discussions.
\end{acknowledgments}

\bibliography{cpv}

\end{document}